\documentclass[pra,aps,showpacs,twocolumn]{revtex4}
\usepackage{amssymb}
\usepackage{amsmath}
\usepackage{mathtools}
\usepackage[dvipdfm]{hyperref}
\usepackage{hyperref}
\allowdisplaybreaks

\begin{document}

\title{Coherence of quantum channels}
\author{Jianwei Xu}
\email{xxujianwei@nwafu.edu.cn}
\affiliation{College of Science, Northwest A\&F University, Yangling, Shaanxi 712100,
China}
\date{\today }

\begin{abstract}
We investigate the coherence of quantum channels and establish a resource
theory for quantifying the coherence of quantum channels via Choi matrix. To
this aim, we define the incoherent channels and incoherent superchannels.
This theory recovers the case of quantum states when we view quantum states
as a special case of quantum channels and also, this theory allows some
analytical expressions for coherence measures.
\end{abstract}

\maketitle



\section{Introduction}

Coherence is a fundamental feature of quantum physics. In recent years,
there have been many papers devoted to study the properties of coherence of
quantum states, such as quantifications, interconvertions, interpretations
and applications ( see reviews \cite{Plenio-2016-RMP,Hu2018-PhysicsReports} and references therein).

All these research for coherence of quantum states extensively enriched our
understanding for quantum theory and leads to many applications. While the
coherence of quantum states is still under active research, recently many
researchers begin to consider the coherence of quantum channels \cite%
{Winter-2017-PRA, Plenio-2019-PRL, LiuYuan-2019, LiuWinter-2019}. The
motivation to study the coherence of channels is somewhat obvious since most
quantum information processings involve and depend on the properties of
quantum channels. Among these research, a natural scheme is often adopted to
define the coherence of a channel as the optimization of coherence over all
output states from this channel. However, such kind of definitions are
generally very hard to calculate even numerically.

In this work, we investigate the coherence of quantum channels from a new
perspective. After some preliminaries about Choi matrices of channel and superchannel (Sec. II), we provide a definition of incoherent channels and then
define and characterize some kinds of superchannels (Sec. III). Especially, we provide
a framework for quantifying the coherence of channels via Choi matrix (Sec. IV). Our
framework unifies the coherence quantification of channels and states in the
sense that states can be viewed as special channels. A conclusion is given in Sec. V.

\section{Preliminaries}

For clarity, we first give some prerequisites and notations, and also we
postpone most proofs of this work to the Appendix part. Let $H^{A}$, $H^{B}$
be two Hilbert spaces with dimensions $|A|,|B|$, and $\{|j\rangle
\}_{j}=\{|k\rangle \}_{k}$, $\{|\alpha \rangle \}_{\alpha }=\{|\beta \rangle
\}_{\beta }$ be orthonormal bases of $H^{A}$, $H^{B}$, respectively. We
always assume the orthonormal bases are fixed, i.e., base dependent, and
adopt the tensor basis $\{|k\rangle |\alpha \rangle \}_{j\alpha }$ as fixed
basis when considering the multipartite system $H^{AB}=H^{A}\otimes H^{B}$.
Let $\mathcal{D}_{A},\mathcal{D}_{B},$ be the set of all density operators
on $H^{A}$ and $H^{B}$ respectively, and $\mathcal{C}_{AB}$ denote the set
of all channels from $\mathcal{D}_{A}$ to $\mathcal{D}_{B}$. A quantum
channel $\phi \in \mathcal{C}_{AB}$ can be represented by the Kraus
operators $\phi =\{M_{m}\}_{m}$ with $M_{m}^{\dagger }M_{m}=I^{A}$ the
identity on $H^{A}$, or by the Choi matrix
\begin{eqnarray}
J_{\phi } &=&\sum_{jk}|j\rangle \langle k|\otimes \phi (|j\rangle \langle k|)
\label{eq1} \\
&=&\sum_{jk\alpha \beta }\phi _{jk,\alpha \beta }|j\rangle \langle k|\otimes
|\alpha \rangle \langle \beta |,  \label{eq2}
\end{eqnarray}%
with $\phi _{jk,\alpha \beta }=\langle \alpha |\phi (|j\rangle \langle
k|)|\beta \rangle .$

It holds that $J_{\phi }\geq 0,$ $\sum_{\alpha }\phi _{jk,\alpha \alpha
}=\delta _{jk},$ and
\begin{equation}
\sum_{n}M_{n\alpha j}M_{n\beta k}^{\ast }=\phi _{jk,\alpha \beta },
\label{eq3}
\end{equation}%
with $M_{m}=\sum_{\alpha j}M_{n\alpha j}|\alpha \rangle \langle j|.$ Note
that $J_{\phi }\geq 0$ means $J_{\phi }$ is positive semidefinite, and $%
\sum_{\alpha }\phi _{jk,\alpha \alpha }=\delta _{jk}$ is equivalent to tr$%
_{B}J_{\phi }=$ $I^{A}$ or $\sum_{m}M_{m}^{\dagger }M_{m}=I^{A}$ or tr$\phi
(|j\rangle \langle k|)=\delta _{jk}.$

We remark that expressions in Eqs. \eqref{eq2}, \eqref{eq3} are used in \cite%
{Plenio-2019-PRL}.

The completely dephsing channel $\Delta ^{A}\in \mathcal{C}_{AA}$ is defined
as
\begin{eqnarray}
\Delta ^{A}(\rho ^{A})=\sum_{j}\langle j|\rho ^{A}|j\rangle |j\rangle
\langle j|, \rho ^{A}\in \mathcal{D}_{A}.
\end{eqnarray}
The notations $\Delta ^{B}\in \mathcal{C}_{BB}$, $I^{A},$ $%
I^{AB}=I^{A}\otimes I^{B},$ are similarly defined, and contractions $%
|j\alpha \rangle =|j\rangle |\alpha \rangle $, $|j\alpha j^{\prime }\alpha
^{\prime }\rangle =|j\rangle |\alpha \rangle |j^{\prime }\rangle |\alpha
^{\prime }\rangle ,$ complex number conjugate $\ast ,$ matrix transpose $t,$
Hermitian conjugate $\dagger $ are used.

When we consider the other systems $A^{\prime}$, $B^{\prime }$, the similar
symbols are defined, such as Hilbert spaces $H^{A^{\prime }}$, $H^{B^{\prime
}},$ dimensions $|A^{\prime }|$, $|B^{\prime }|,$ fixed orthonormal bases $%
\{|j^{\prime }\rangle \}_{j^{\prime }}=\{|k^{\prime }\rangle \}_{k^{\prime
}} $, $\{|\alpha ^{\prime }\rangle \}_{\alpha ^{\prime }}=\{|\beta ^{\prime
}\rangle \}_{\beta ^{\prime }}$ of $H^{A^{\prime }}$, $H^{B^{\prime }},$ the
sets of density operators $\mathcal{D}_{A^{\prime }},\mathcal{D}_{B^{\prime
}},$ the set of quantum channels $\mathcal{C}_{A^{\prime }B^{\prime }}$, the
identity $I^{A^{\prime }},$ etc.

A superchannel $\Theta \in \mathcal{SC}_{ABA^{\prime }B^{\prime }}$, is a
linear map from $\mathcal{C}_{AB}$ to $\mathcal{C}_{A^{\prime }B^{\prime }},$
we express its Choi matrix as
\begin{eqnarray}
J_{\Theta } &=&\sum_{jk\alpha \beta }|j\alpha \rangle \langle k\beta
|\otimes \Theta (|j\alpha \rangle \langle k\beta |)  \label{eq5} \\
&=&\sum_{jk\alpha \beta }\Theta _{jk,\alpha \beta ,j^{\prime }k^{\prime
},\alpha ^{\prime }\beta ^{\prime }}|j\alpha j^{\prime }\alpha ^{\prime
}\rangle \langle k\beta k^{\prime }\beta ^{\prime }|,  \label{eq6}
\end{eqnarray}%
with $\Theta _{jk,\alpha \beta ,j^{\prime }k^{\prime },\alpha ^{\prime
}\beta ^{\prime }}=\langle j^{\prime }\alpha ^{\prime }|\Theta (|j\alpha
\rangle \langle k\beta |)|k^{\prime }\beta ^{\prime }\rangle .$ The
expression in Eq. \eqref{eq6} is just because of the linearity of
superchannel. It is worth emphasizing that there is an analogy between Eq. %
\eqref{eq1} and Eq. \eqref{eq5}, as well as between Eq. \eqref{eq2} and Eq. %
\eqref{eq6}, when we regard the bipartite states $\{|j\alpha \rangle
\}_{j\alpha }$ as single partite states.

A superchannel should satisfy some conditions \cite{Chiribella-2008-EPL,
Gour-2019-IEEE}, when we express the Choi matrix $J_{\Theta }$ in terms of $%
\Theta _{jk,\alpha \beta ,j^{\prime }k^{\prime },\alpha ^{\prime }\beta
^{\prime }}$ above, we have Proposition 1 below.

$\boldsymbol{Proposition \ 1.}$ $\Theta \in \mathcal{SC%
}_{ABA^{\prime }B^{\prime }}$ iff (if and only if) $J_{\Theta }\geq 0,$ and
\begin{eqnarray}
\sum_{\alpha ^{\prime }}\Theta _{jk,\alpha \beta ,j^{\prime }k^{\prime
},\alpha ^{\prime }\alpha ^{\prime }}&=&\delta _{\alpha \beta }\rho
_{j^{\prime }k^{\prime }}^{(jk)},  \label{eq7} \\
\sum_{j}\rho _{j^{\prime }k^{\prime }}^{(jj)}&=&\delta _{j^{\prime
}k^{\prime }}.  \label{eq8}
\end{eqnarray}

Similar to the case of channel, a superchannel $\Theta $ also has the
expression of Kraus operators $\Theta =\{\mathcal{M}_{m}\}_{m}\in \mathcal{SC%
}_{ABA^{\prime }B^{\prime }},$ with
\begin{eqnarray}
J_{\Theta (\phi )}&=&\sum_{m}\mathcal{M}_{m}J_{\phi }\mathcal{M}%
_{m}^{\dagger }, \forall \ \phi \in \mathcal{C}_{AB},  \label{eq9} \\
\Theta _{jk,\alpha \beta ,j^{\prime }k^{\prime },\alpha ^{\prime }\beta
^{\prime }}&=&\sum_{m}\mathcal{M}_{mj^{\prime }j,\alpha ^{\prime }\alpha }%
\mathcal{M}_{mk^{\prime }k,\beta ^{\prime }\beta }^{\ast }.  \label{eq10}
\end{eqnarray}
where
\begin{eqnarray}
\mathcal{M}_{m}=\sum_{j^{\prime }j,\alpha ^{\prime }\alpha }\mathcal{M}%
_{mj^{\prime }j,\alpha ^{\prime }\alpha }|j^{\prime }\alpha ^{\prime
}\rangle \langle j\alpha |.
\end{eqnarray}

Superchannel can be physically implemented as a simple quantum circuit \cite%
{Chiribella-2008-EPL, Gour-2019-IEEE}.

\section{Incoherent channels}

We first specify the incoherent channels. Recall that for quantum states, a
state $\sigma ^{A}\in \mathcal{D}_{A}$ is called incoherent if
\begin{equation}
\Delta ^{A}(\sigma ^{A})=\sigma ^{A}.
\end{equation}%
$\Delta ^{A}$ is a resource destroying map \cite{Lloyd-2017-PRL}, that is to
say, $\Delta ^{A}(\sigma ^{A})=\sigma ^{A}$ for any incoherent state $\sigma
^{A}\in \mathcal{D}_{A}$ and $\Delta ^{A}(\rho ^{A})$ is incoherent for any $%
\rho ^{A}\in \mathcal{D}_{A}.$ For channels, we give the following
definition.

$\boldsymbol{Definition \ 1.}$ We call a channel $\phi \in \mathcal{C}_{AB}$
an incoherent channel (IC) if
\begin{eqnarray}
\Upsilon (\phi )=\phi,
\end{eqnarray}
where
\begin{eqnarray}
\Upsilon (\phi )=\Delta ^{B}\phi \Delta ^{A}, \forall \ \phi \in \mathcal{C}%
_{AB}.
\end{eqnarray}

We denote the set of all incoherent channels in $\mathcal{C}_{AB}$ as $%
\mathcal{IC}_{AB},$ and the set of all incoherent channels in $\mathcal{C}%
_{A^{\prime }B^{\prime }}$ as $\mathcal{IC}_{A^{\prime }B^{\prime }}.$ The
IC defined here is also called classical channel \cite{Gour-note}.

It is easy to check that $\Upsilon $ also is a resource destroying map \cite%
{Lloyd-2017-PRL}, i.e., $\Upsilon (\phi )=\phi $ for any incoherent channel $%
\phi \in \mathcal{C}_{AB}$ and $\Upsilon (\phi )$ is incoherent for any $%
\phi \in \mathcal{C}_{AB}.$ We call $\Upsilon $ the completely dephasing
superchannel.

$\boldsymbol{Proposition \ 2.}$ $\phi \in \mathcal{C}_{AB},$ then $\phi \in
\mathcal{IC}_{AB}$ iff
\begin{eqnarray}
\phi _{jk,\alpha \beta }=\delta _{jk}\delta _{\alpha \beta }\phi _{jj,\alpha
\alpha }, \forall \ j,k,\alpha ,\beta ,  \label{eq15}
\end{eqnarray}
that is, $J_{\phi }$ is diagonal.

We see that the intuitive explanation of incoherent channels is, for such
channels quantum coherence of a state (off-diagonal entries) can be neither
input into nor output from them.

It can be checked that the channel $\chi \in \mathcal{IC}_{AB}$ admits the
Kraus operators $\chi =\{M_{\alpha j}\}_{\alpha j}$ with $M_{\alpha j}=\sqrt{%
\chi _{jj,\alpha \alpha }}|\alpha \rangle \langle j|.$

It is easy to see that $\mathcal{IC}_{AB}$ is a convex set, Theorem 3 below
characterizes the structure of $\mathcal{IC}_{AB}.$

$\boldsymbol{Theorem \ 3.}$
\begin{eqnarray}
(1). \ \mathcal{IC}_{AB} =conv\{\phi \in \mathcal{C}_{AB}|\phi _{jk,\alpha
\beta }=\delta _{jk}\delta _{\alpha \beta }\delta _{\alpha ,f(j)}\}, \ \
\label{eq16} \\
(2). \ \mathcal{IC}_{AB} \subsetneqq \mathcal{PIO}_{AB}, \ \ \ \ \ \ \ \ \ \
\ \ \ \ \ \ \ \ \ \ \ \ \ \ \ \ \ \ \ \ \ \ \ \ \ \ \ \ \ \ \   \label{eq17}
\end{eqnarray}%
where $conv$ means convex hull, $f(j)\in \{\alpha \}_{\alpha =1}^{|B|},%
\mathcal{PIO}_{AB}$ is the set of all PIOs (physically incoherent
operations) \cite{ChitambarGour-2016-PRL, ChitambarGour-2016-PRA}.

For the free operations of incoherent channels, we propose the definitions
of MISC and ISC, they correspond to MIO (maximally incoherent operation)
\cite{Aberg-2016-arxiv} and IO (incoherent operation) \cite%
{BCP-2014-PRL,Guo-2015-PRA} of quantum states.

$\boldsymbol{Definition \ 2.}$ A superchannel $\Theta $ is called a
maximally incoherent superchannel (MISC) if $\Theta (\chi )\in \mathcal{IC}%
_{A^{\prime }B^{\prime }}$ for $\forall $ $\chi \in \mathcal{IC}_{AB}.$

We denote the set of all MISCs by $\mathcal{MISC}_{ABA^{\prime }B^{\prime
}}. $

$\boldsymbol{Definition\ 3.}$ A superchannel $\Theta \in \mathcal{SC}%
_{ABA^{\prime }B^{\prime }}$ is called an incoherent superchannel (ISC) if
it admits an expression of Kraus operators $\Theta =\{\mathcal{M}_{m}\}_{m}$
such that for each $m$,
\begin{equation}
\mathcal{M}_{m}=\sum_{j\alpha }\mathcal{M}_{mj\alpha }|f(j\alpha )\rangle
\langle j\alpha |,  \label{eq18}
\end{equation}%
where $f(j\alpha )=f(j,\alpha )\in \{(j^{\prime },\alpha ^{\prime
})|_{j^{\prime }=1}^{|A^{\prime }|},_{\alpha ^{\prime }=1}^{|B^{\prime
}|}\}. $ We call $\{\mathcal{M}_{m}\}_{m}$ an incoherent expression for ISC $%
\Theta $ if for each $m,$ $\mathcal{M}_{m}$ has the form in Eq. \eqref{eq18}.

We denote the set of all ISCs by $\mathcal{ISC}_{ABA^{\prime }B^{\prime }}.$

It can be checked that $\mathcal{M}_{m}J_{\chi }\mathcal{M}_{m}^{\dagger }$
is diagonal for any $\mathcal{M}_{m}$ in Eq. \eqref{eq18} and any $\chi \in
\mathcal{IC}_{AB}.$

It follows that
\begin{equation}
\mathcal{ISC}_{ABA^{\prime }B^{\prime }}\subsetneqq \mathcal{MISC}%
_{ABA^{\prime }B^{\prime }}\subsetneqq \mathcal{SC}_{ABA^{\prime }B^{\prime
}},
\end{equation}%
and they are all convex sets.

Unitary channels are very fundamental in the sense of Stinespring dilation.
For superchannels, we give the definition of preunitary superchannel.

$\boldsymbol{Definition \ 4.}$ For $|A|\geq |A^{\prime }|$ and $|B|\leq
|B^{\prime }|,$ a superchannel $\Theta =\{\mathcal{U}\}\in \mathcal{SC}%
_{ABA^{\prime }B^{\prime }}$ is called preunitary if it has an expression of
only one Kraus operator $\mathcal{U}.$ When $|A|=|A^{\prime }|$ and $%
|B|=|B^{\prime }|,$ a preunitary superchannel is called a unitary
superchannel, for such case $\mathcal{U}^{\dagger }\mathcal{U=UU}^{\dagger
}=I^{AB}.$

Unitary incoherent channels are important for the discussion of PIO \cite%
{ChitambarGour-2016-PRL, ChitambarGour-2016-PRA}. For the case of
superchannel, we have following theorem.

$\boldsymbol{Theorem \ 4.}$ For $|A|\geq |A^{\prime }|$ and $|B|\leq
|B^{\prime }|,$ a superchannel $\Theta =\{\mathcal{U}\}\in \mathcal{SC}%
_{ABA^{\prime }B^{\prime }}$ is preunitary, iff it has the form
\begin{eqnarray}
\mathcal{U}=U\otimes V,
\end{eqnarray}
where $U$ is a $|A^{\prime } |\times |A|$ coisometry, i.e., $UU^{\dagger
}=I^{A^{\prime }}$, $V$ is a $|B^{\prime } |\times |B|$ isometry, i.e., $%
V^{\dagger }V=I^{B}.$ Further, if $\Theta =\{\mathcal{U}\}$ is preunitary
and incoherent, then $U$ and $V$ all have at most one nonzero element in
each column.

\section{A framework for quantifying coherence of quantum channels}

Quantum resource theory (QRT) provides a powerful tool for quantifying a
certain quantum feature possessed in quantum systems or quantum processes
(see \cite{ChitambarGour-2019-RMP} and references therein). In QRT, the
specification of free operations is in principle not unique and different
settings are motivated by various considerations. Recently a QRT for
coherence of channels has been proposed in \cite{Plenio-2019-PRL} where free
channels are detection incoherent channels or creation incoherent channels.
In this section, with $\mathcal{IC}$ as free channels and $\mathcal{ISC}$ as
free superchannels, and inspired by the BCP framework for quantifying the
coherence of quantum states in \cite{BCP-2014-PRL}, we establish a framework
for quantifying the coherence of quantum channels.

We propose the necessary conditions that any coherence measure $C$ for
quantum channels should satisfy.

(C1). Faithfulness: $C(\phi )\geq 0$ for any $\phi \in \mathcal{C}_{AB},$
and $C(\phi )=0$ iff $\phi \in \mathcal{IC}_{AB}.$

(C2a). Nonincreasing under ISC: $C(\phi )\geq C[\Theta (\phi )]$ for any $%
\Theta \in \mathcal{ISC}_{ABA^{\prime }B^{\prime }}.$

(C2b). Nonincreasing under ISC on average: $C(\phi )\geq \sum_{m}p_{m}C(\phi
_{m})$ for any $\Theta \in \mathcal{ISC}_{ABA^{\prime }B^{\prime }}$, with $%
\{\mathcal{M}_{m}\}_{m}$ an incoherent expression of $\Theta $, $p_{m}=\frac{%
tr(\mathcal{M}_{m}J_{\phi }\mathcal{M}_{m}^{\dagger })}{|A^{\prime }|}$, $%
J_{\phi _{m}}=|A^{\prime }|\frac{\mathcal{M}_{m}J_{\phi }\mathcal{M}%
_{m}^{\dagger }}{tr(\mathcal{M}_{m}J_{\phi }\mathcal{M}_{m}^{\dagger })}.$

(C3). Convexity: $C(\sum_{m}p_{m}\phi _{m})\leq \sum_{m}p_{m}C(\phi _{m}),$
for any $\{\phi _{m}\}_{m}\subset \mathcal{C}_{AB}$ and probability $%
\{p_{m}\}_{m}.$

Notice that (C3) and (C2b) automatically imply (C2a).

Note that in (C2b), $J_{\phi _{m}}=|A^{\prime }|\frac{\mathcal{M}_{m}J_{\phi
}\mathcal{M}_{m}^{\dagger }}{tr(\mathcal{M}_{m}J_{\phi }\mathcal{M}%
_{m}^{\dagger })}$ is not necessarily a Choi matrix for quantum channel, so
this framework implies that any coherence measure $C$ should properly
defined on such $J_{\phi _{m}}.$

From now on we discuss some properties of this framework. One advantage of
this framework is that we can get some results for quantum channels by using
the known results of quantum states.

$\boldsymbol{Theorem \ 5.}$ If $C$ is a coherence measure for quantum states
in the BCP framework \cite{BCP-2014-PRL}, then
\begin{eqnarray}
C(\phi )=C(\frac{J_{\phi }}{|A|}), \phi \in \mathcal{C}_{AB},
\end{eqnarray}
is a coherence measure for quantum channels.

The proof is straightforward by checking the four conditions above.

From Theorem 5 we can get many coherence measures for channels corresponding
to the coherence measures for quantum states, for examples the coherence
measure based on $l_{1}$ norm $C_{l_{1}}$ \cite{BCP-2014-PRL}, coherence
measure based on relative
entropy $C_{r}$ \cite{BCP-2014-PRL}, based on Tsallis relative entropy $C_{\alpha }$
\cite{Yu-2017-PRA, Yu-2018-SR}, based on robustness $C_{R}$ \cite{Adesso-2016-PRL,
Adesso-2016-PRA}, based on trace norm $C_{tr}$ \cite{Tong-2016-PRA}, and the geometric coherence \cite{Streltsov-2015-PRL}.

$\boldsymbol{Definition\ 5.}$ A channel is called a channel with maximal
coherence, if it reaches the maximum for any coherence measure of channels.

$\boldsymbol{Theorem\ 6.}$ For $\phi \in $ $\mathcal{C}_{AB}$, if $\frac{%
J_{\phi }}{|A|}$ is a quantum state with maximal coherence on the composite
space $H^{A}\otimes H^{B}$, then $\phi $ is a channel with maximal
coherence. As a result, for $|A|\leq |B|,$ the isometry channel $U_{\max
}\in $ $\mathcal{C}_{AB},$
\begin{eqnarray}
U_{\max }(|j\rangle )=\frac{1}{\sqrt{|B|}}\sum_{\alpha =1}^{|B|}e^{i\theta
_{j\alpha }}|\alpha \rangle ,  \label{eq22} \\
\frac{1}{|B|}\sum_{\alpha =1}^{|B|}e^{i(\theta _{j\alpha }-\theta _{k\alpha
})}=\delta _{jk},  \label{eq23}
\end{eqnarray}
is a channel with maximal coherence. Especially for $|A|\leq |B|,$ the
isometry channel
\begin{eqnarray}
U_{\max }^{(0)}(|j\rangle )=\frac{1}{\sqrt{|B|}}\sum_{\alpha =1}^{|B|}e^{%
\frac{2\pi i}{|B|}j\alpha }|\alpha \rangle ,
\end{eqnarray}
is a channel with maximal coherence.

\emph{Proof.} The first statement that $\phi $ is a channel with maximal coherence
is a result of Theorem 5. From Eqs. \eqref{eq22} and \eqref{eq23} we can
check that $\frac{J_{U_{\max }}}{|A|}$ is a quantum state with maximal
coherence \cite{Fan-2016-PRA} and $U_{\max }$ is an isometry channel. $\Box$

Note that if $\rho ^{AB}$ is a quantum state with maximal coherence on $%
H^{AB},$ then $|A|\rho ^{AB}$ is not necessarily a Choi matrix for any
channel. For example,
\begin{eqnarray}
|\psi ^{AB}\rangle =\frac{1}{\sqrt{|A||B|}}\sum_{j=1}^{|A|}\sum_{\alpha
=1}^{|B|}|j\rangle |\alpha \rangle ,
\end{eqnarray}
is a quantum state with maximal coherence on $H^{AB}$ \cite{BCP-2014-PRL},
but $|A||\psi ^{AB}\rangle \langle \psi ^{AB}|$ is not a Choi matrix for any
channel since tr$_{B}(|A||\psi ^{AB}\rangle \langle \psi ^{AB}|)\neq I^{A}$
for $|A|\geq 2.$

Another advantage of this framework is that we can regard it as a unified
theory for quantifying coherence of both quantum states and quantum
channels. A quantum state of system A can be viewed as a special quantum
channel in $\mathcal{C}_{AB}$ for $|A|=1.$ For $|A|=|A^{\prime }|=1,$ we see
that MISC degenerates to MIO, ISC degenerates to IO, and preunitary
superchannel degenerates to isometry channel. With these observations we
have proposition 7 below.

$\boldsymbol{Proposition \ 7.}$ A coherence measure for quantum channels
degenerates to a coherence measure for quantum states in BCP framework \cite%
{BCP-2014-PRL} when $|A|=|A^{\prime }|=1.$

For the case $|B|=1,$ there is only one channel $\phi \in $ $\mathcal{C}%
_{AB} $ with Choi matrix $J_{\phi }=I^{A},$ which is the action of trace $%
\phi (\rho ^{A})=tr(\rho ^{A})=1.$ Note that this is an incoherent channel,
this fact coincides with our intuition.

The third advantage of this framework is that it possesses the monotonicity
under composition below.

$\boldsymbol{Proposition \ 8.}$ Monotonicity under composition.

(1). For $\phi \in \mathcal{C}_{AB}$, $\chi \in \mathcal{IC}_{BB^{\prime }}$%
,
\begin{eqnarray}
C(\chi \circ \phi )\leq C(\phi ).
\end{eqnarray}
Especially, for $\phi \in \mathcal{C}_{AB}$, $\Delta ^{B}\in \mathcal{IC}%
_{BB}$,
\begin{eqnarray}
C(\Delta ^{B}\circ \phi )\leq C(\phi ).
\end{eqnarray}
(2). For $|A^{\prime }|\leq|A|$, $\phi \in \mathcal{C}_{AB}$, $\chi \in
\mathcal{IC}_{A^{\prime }A}$, $\sum_{j^{\prime }}\chi _{j^{\prime }j^{\prime
},jj}\leq 1, \forall \ j,$
\begin{eqnarray}
C(\phi \circ \chi )\leq C(\phi ).
\end{eqnarray}
Especially, for $\phi \in \mathcal{C}_{AB}$, $\Delta ^{A}\in \mathcal{IC}%
_{AA}$,
\begin{eqnarray}
C(\phi \circ \Delta ^{A})\leq C(\phi ).
\end{eqnarray}

Finally we give an example to show the unity of quantifying coherence for
both channels and states under Theorem 5.

\emph{Example.} Consider the channel $\widetilde{\phi }\in $ $\mathcal{C}%
_{AB},$
\begin{equation}
\widetilde{\phi }(\rho ^{A})=p\phi (\rho ^{A})+(1-p)\rho ^{B}, \forall \
\rho ^{A}\in \mathcal{D}_{A},
\end{equation}
with $p\in \lbrack 0,1],$ $\phi \in $ $\mathcal{C}_{AB}$, $\phi $ is a MIO,
and fixed $\rho ^{B}\in \mathcal{D}_{B}.$

Since $\phi $ is a MIO then
\begin{equation}
\phi _{jj,\alpha \beta }=\delta _{\alpha \beta }\phi _{jj,\alpha \alpha }.
\end{equation}

The channel
\begin{equation}
\phi_{1} (\rho ^{A})=\rho ^{B}, \forall \ \rho ^{A}\in \mathcal{D}_{A},
\end{equation}
has the Choi matrix
\begin{equation}
J_{\phi_{1}}=I^{A}\otimes\rho ^{B}.
\end{equation}

Hence we can equivalently express $\widetilde{\phi }$ as
\begin{equation}
\widetilde{\phi }_{jk,\alpha \beta }=p\phi _{jk,\alpha \beta }+(1-p)\delta
_{jk}\rho _{\alpha \beta }^{B},
\end{equation}%
with $\rho _{\alpha \beta }^{B}=\langle\alpha|\rho^{B}|\beta\rangle.$

Using the coherence measure $C_{l_{1}},$
\begin{equation}
C_{l_{1}}(\phi)=\frac{1}{|A|}\sum _{jk \alpha \beta } |\phi _{jk,\alpha
\beta }|-1, \forall \ \phi\in \mathcal{C}_{AB},
\end{equation}%
we get
\begin{equation}
C_{l_{1}}(\widetilde{\phi })=pC_{l_{1}}(\phi )+(1-p)C_{l_{1}}(\rho ^{B}).
\end{equation}%
We see that under the coherence measure $C_{l_{1}}$, the total channel
coherence $C_{l_{1}}(\widetilde{\phi })$ includes two parts, $C_{l_{1}}(\phi
)$ accounts for the contribution of channel $\phi $, and $C_{l_{1}}(\rho
^{B})$ accounts for the contribution of state $\rho ^{B}.$

\section{Conclusion and outlook}

We investigated the coherence of quantum channels, defined IC, MISC, ISC,
preunitary superchannel, and under quantum resource theory we established a
framework to quantify the coherence of quantum channels via Choi matrix. We
showed that this framework has many advantages, such as getting some results
from coherence theory of quantum states, allowing for a unified viewpoint
combining coherence of channels and states, and satisfying a monotonicity of
composition. We hope this work will open a new way to explore the coherence
of quantum channels.

There are many questions for future research after this work. One is that
whether or not there exist two or more coherence measures for channels they
degenerate to the same coherence measure for quantum states when $%
|A|=|A^{\prime }|=1.$ If it is not true, then coherence measures for both
channels and states are the same under Theorem 5.

Another important question is the physical interpretations for the coherence
measures of channels. Many works has been done for some coherence measures
of quantum states such as \cite{Yang-2016-PRL,Wu-2017-PRL} which may provide
inspiring evidences.

Also, Gaussian channels are very important in both theories and experiments,
while it seems difficult to address the coherence since coherence is
primarily defined on orthonormal bases but Gaussian states and Gaussian
channels intrinsically defined in phase space.

\section*{ACKNOWLEDGMENTS}

The author thanks Gilad Gour, Carlo Maria Scandolo, Gaurav Saxena, Yunlong
Xiao and Si-Ren Yang for helpful discussions. This work is supported by the
China Scholarship Council (CSC, No. 201806305050).

\section*{Appendix}

In this appendix, we provide some necessary details and proofs for the
results in the main text. \setcounter{equation}{0} \renewcommand%
\theequation{A\arabic{equation}}

\subsection{ Choi matrices of Channels and superchannels}

The Choi matrix $J_{\phi }$ of a channel $\phi \in \mathcal{C}_{AB}$ is
defined in Eq. \eqref{eq1}. Choi theorem says $\phi $ is completely positive
definite iff $J_{\phi }\geq 0.$ Since $J_{\phi }\geq 0$ then $J_{\phi }$ has
the decomposition of the form
\begin{eqnarray}
J_{\phi } &=&\sum_{m}|M_{m}\rangle \langle M_{m}|  \notag \\
&=&\sum_{m}(\sum_{\alpha j}M_{m\alpha j}|j\alpha \rangle )(\sum_{k\beta
}M_{m\beta k}^{\ast }\langle k\beta |)  \notag \\
&=&\sum_{mjk\alpha \beta }M_{m\alpha j}M_{m\beta k}^{\ast }|j\rangle \langle
k|\otimes |\alpha \rangle \langle \beta |  \notag \\
&=&\sum_{jk}|j\rangle \langle k|\otimes \sum_{m}M_{m}|j\rangle \langle
k|M_{m}^{\dagger },
\end{eqnarray}
where $|M_{m}\rangle \in H^{AB},$
\begin{eqnarray}
|M_{m}\rangle &=&\sum_{j\alpha }M_{m\alpha j}|j\alpha \rangle , \\
M_{m} &=&\sum_{\alpha j}M_{m\alpha j}|\alpha \rangle \langle j|.
\end{eqnarray}%
Compare to Eq. \eqref{eq1} we see that $\phi =\{M_{m}\}_{m}$ and Eq. %
\eqref{eq3} holds.

Similarly, the Choi matrix $J_{\Theta }$ of a superchannel $\Theta \in
\mathcal{SC}_{ABA^{\prime }B^{\prime }}$ is defined in Eq. \eqref{eq5}. Choi
theorem says $\Theta $ is completely positive definite iff $J_{\Theta }\geq
0.$ Since $J_{\Theta }\geq 0$ then $J_{\Theta }$ has the decomposition of
the form
\begin{eqnarray}
J_{\Theta } &=&\sum_{m}|\mathcal{M}_{m}\rangle \langle \mathcal{M}_{m}|
\notag \\
&=&\sum_{m}(\sum_{j\alpha j^{\prime }\alpha ^{\prime }}\mathcal{M}%
_{mj^{\prime }j,\alpha ^{\prime }\alpha }|j\alpha j^{\prime }\alpha ^{\prime
}\rangle )  \notag \\
&& \ \ \ \ \cdot(\sum_{k\beta k^{\prime }\beta ^{\prime }}\mathcal{M}%
_{mk^{\prime }k,\beta ^{\prime }\beta }^{\ast }\langle k\beta k^{\prime
}\beta ^{\prime }|)  \notag \\
&=&\sum_{mjk\alpha \beta j^{\prime }k^{\prime }\alpha ^{\prime }\beta
^{\prime }}\mathcal{M}_{mj^{\prime }j,\alpha ^{\prime }\alpha }\mathcal{M}%
_{mk^{\prime }k,\beta ^{\prime }\beta }^{\ast }  \notag \\
&& \ \ \ \ \ \ \ \ \ \ \ \ \ \ \ \ \ \ \cdot |j\alpha \rangle \langle k\beta
|\otimes |j^{\prime }\alpha ^{\prime }\rangle \langle k^{\prime }\beta
^{\prime }|  \notag \\
&=&\sum_{jk\alpha \beta }|j\alpha \rangle \langle k\beta |\otimes \sum_{m}%
\mathcal{M}_{m}|j\alpha \rangle \langle k\beta |\mathcal{M}_{m}^{\dagger },
\end{eqnarray}%
where $|\mathcal{M}_{m}\rangle \in H^{ABA^{\prime }B^{\prime
}}=H^{AB}\otimes H^{A^{\prime }B^{\prime }},$
\begin{eqnarray}
|\mathcal{M}_{m}\rangle &=&\sum_{j\alpha j^{\prime }\alpha ^{\prime }}%
\mathcal{M}_{mj^{\prime }j,\alpha ^{\prime }\alpha }|j\alpha j^{\prime
}\alpha ^{\prime }\rangle , \\
\mathcal{M}_{m} &=&\sum_{j\alpha j^{\prime }\alpha ^{\prime }}\mathcal{M}%
_{mj^{\prime }j,\alpha ^{\prime }\alpha }|j^{\prime }\alpha ^{\prime
}\rangle \langle j\alpha |.
\end{eqnarray}%
Then we can get Eq. \eqref{eq10}.

It can be checked that \cite%
{Chiribella-2008-EPL,Matthews-2015-IEEE,Gour-2019-IEEE}
\begin{equation}
\phi (\rho ^{A})=tr_{A}[J_{\phi }(\rho ^{A})^{t}\otimes I^{B}],
\end{equation}%
this can be equivalently written as
\begin{equation}
\lbrack \phi (\rho ^{A})]_{\alpha \beta }=\sum_{jk}\rho _{jk}^{A}\phi
_{jk,\alpha \beta },
\end{equation}%
with $\rho _{jk}^{A}=\langle j|\rho^{A}|k\rangle$ and $\lbrack \phi (\rho
^{A})]_{\alpha \beta }=\langle \alpha|\phi (\rho ^{A})|\beta\rangle.$ Since
the channel $\phi $ is trace preserving, then
\begin{equation}
\sum_{\alpha }\phi _{jk,\alpha \alpha }=\delta _{jk}.
\end{equation}

For superchannel $\Theta \in \mathcal{SC}_{ABA^{\prime }B^{\prime }}$, and
channel $\phi \in \mathcal{C}_{AB}$, it holds that \cite%
{Chiribella-2008-EPL, Matthews-2015-IEEE, Gour-2019-IEEE}
\begin{equation}
J_{\Theta (\phi )}=tr_{AB}[J_{\Theta }(J_{\phi })^{t}\otimes I^{A^{\prime
}B^{\prime }}],
\end{equation}%
this can be equivalently expressed as
\begin{equation}
\lbrack \Theta (\phi )]_{j^{\prime }k^{\prime },\alpha ^{\prime }\beta
^{\prime }}=\sum_{jk,\alpha \beta }\phi _{jk,\alpha \beta }\Theta
_{jk,\alpha \beta ,j^{\prime }k^{\prime },\alpha ^{\prime }\beta ^{\prime }}.
\label{eqA11}
\end{equation}%
Taking Eq. \eqref{eq10} into Eq. \eqref{eqA11} we get Eq. \eqref{eq9}.

For $\phi \in \mathcal{C}_{AB}$, $\psi \in \mathcal{C}_{BB^{\prime }}$, the
composition $\psi \circ \phi \in \mathcal{C}_{AB^{\prime }}$, and
\begin{equation}
(\psi \circ \phi )_{jk,\alpha ^{\prime }\beta ^{\prime }}=\sum_{\alpha \beta
}\phi _{ij,\alpha \beta }\psi _{\alpha \beta ,\alpha ^{\prime }\beta
^{\prime }},
\end{equation}
this is equivalent to Eq. (3) in \cite{Chiribella-2008-PRL}.

\subsection{Proof of Proposition 1}

Proposition 1 is equivalent to Eq. (19) in \cite{Gour-2019-IEEE} and also
the result in \cite{Chiribella-2008-EPL}, here we give a proof for being
self contained and consistent with the notations used in this work.

From Choi theorem, $\Theta $ is completely positive iff $J_{\Theta }\geq 0.$
Suppose $\phi \in \mathcal{C}_{AB}$, then $J_{\phi }\geq 0$ (completely
positive) and tr$\phi (|j\rangle \langle k|)=\delta _{jk}$ (trace
preserving). Note that $J_{\phi }\geq 0$ implies $J_{\phi }^{\dagger
}=J_{\phi },$ $J_{\Theta }\geq 0$ implies $J_{\Theta }^{\dagger }=J_{\Theta
},$ that is
\begin{eqnarray}
\phi _{jk,\alpha \beta }&=&\phi _{kj,\beta \alpha }^{\ast }, \\
\Theta _{jk,\alpha \beta ,j^{\prime }k^{\prime },\alpha ^{\prime }\beta
^{\prime }}&=&\Theta _{kj,\beta \alpha ,k^{\prime }j^{\prime },\beta
^{\prime }\alpha ^{\prime }}^{\ast }.
\end{eqnarray}
If $\Theta (\phi )$ is trace-preserving, we have
\begin{eqnarray}
\sum_{jk,\alpha \beta ,\alpha ^{\prime }}\phi _{jk,\alpha \beta }\Theta
_{jk,\alpha \beta ,j^{\prime }k^{\prime },\alpha ^{\prime }\alpha ^{\prime
}}=\delta _{j^{\prime }k^{\prime }}, \forall \ \phi \in \mathcal{C}_{AB}. \
\   \label{eqA15}
\end{eqnarray}
For clarity, we set four steps to complete this proof.

Step 1. Let $\phi _{jk,\alpha \beta }=\delta _{jk}\delta _{\alpha \beta
}\phi _{jj,\alpha \alpha },$ $\phi _{jj,\alpha \alpha }\geq 0,$ $%
\sum_{\alpha }\phi _{jj,\alpha \alpha }=1,$ then
\begin{eqnarray}
\sum_{j\alpha \alpha ^{\prime }}\phi _{jj,\alpha \alpha }\Theta _{jj,\alpha
\alpha ,j^{\prime }k^{\prime },\alpha ^{\prime }\alpha ^{\prime }}=\delta
_{j^{\prime }k^{\prime }}.  \label{eqA16}
\end{eqnarray}
Varying $\{\phi _{jj,\alpha \alpha }\}_{j\alpha }$ on the domain of $\{\phi
_{jj,\alpha \alpha }\in R: \phi _{jj,\alpha \alpha }\geq 0$, $\sum_{\alpha
}\phi _{jj,\alpha \alpha }=1 \}$, we get
\begin{eqnarray}
\sum_{\alpha ^{\prime }}\Theta _{jj,\alpha \alpha ,j^{\prime }k^{\prime
},\alpha ^{\prime }\alpha ^{\prime }}=:\rho _{j^{\prime }k^{\prime }}^{(jj)},
\label{eqA17}
\end{eqnarray}
independent of $\alpha ,$ and
\begin{eqnarray}
&&\sum_{j\alpha \alpha ^{\prime }}\phi _{jj,\alpha \alpha }\Theta
_{jj,\alpha \alpha ,j^{\prime }k^{\prime },\alpha ^{\prime }\alpha ^{\prime
}}  \notag \\
&=&\sum_{j}(\sum_{\alpha }\phi _{jj,\alpha \alpha })(\sum_{\alpha ^{\prime
}}\Theta _{jj,\alpha \alpha ,j^{\prime }k^{\prime },\alpha ^{\prime }\alpha
^{\prime }})  \notag \\
&=&\sum_{j}\rho _{j^{\prime }k^{\prime }}^{(jj)}=\delta _{j^{\prime
}k^{\prime }}.
\end{eqnarray}
This proves Eq. \eqref{eq8}.

Step 2. Fix $j_{0},k_{0}\in \{j\}_{j=1}^{|A|},j_{0}<k_{0},$ let
\begin{eqnarray}
\phi _{jj,\alpha \alpha }=\frac{1}{|B|}, \forall \ j,\alpha , \\
\phi _{j_{0}k_{0},\alpha \alpha }=\phi _{k_{0}j_{0},\alpha \alpha }^{\ast },
\forall \ \alpha ,
\end{eqnarray}
and others $\phi _{jk,\alpha \beta }=0.$ We also let $|\phi
_{j_{0}k_{0},\alpha \alpha }|\leq \frac{1}{|B|}$ for $\forall \ \alpha $ to
ensure $J_{\phi }\geq 0$ according to the Gersgorin discs theorem.

For this case, subtract \eqref{eqA16} from \eqref{eqA15} we get
\begin{eqnarray}
\sum_{\alpha \alpha ^{\prime }}(\phi _{j_{0}k_{0},\alpha \alpha }\Theta
_{j_{0}k_{0},\alpha \alpha ,j^{\prime }k^{\prime },\alpha ^{\prime }\alpha
^{\prime }} \ \ \ \ \ \ \ \ \ \   \notag \\
+\phi _{k_{0}j_{0},\alpha \alpha }\Theta _{k_{0}j_{0},\alpha \alpha
,j^{\prime }k^{\prime },\alpha ^{\prime }\alpha ^{\prime }})=0,
\end{eqnarray}
\begin{eqnarray}
\sum_{\alpha }\phi _{j_{0}k_{0},\alpha \alpha }\sum_{\alpha ^{\prime
}}\Theta _{j_{0}k_{0},\alpha \alpha ,j^{\prime }k^{\prime },\alpha ^{\prime
}\alpha ^{\prime }} \ \ \ \ \ \   \notag \\
+\sum_{\alpha }\phi _{j_{0}k_{0},\alpha \alpha }^{\ast }\sum_{\alpha
^{\prime }}\Theta _{j_{0}k_{0},\alpha \alpha ,k^{\prime }j^{\prime },\alpha
^{\prime }\alpha ^{\prime }}^{\ast } =0,
\end{eqnarray}
Notice that $\sum_{\alpha }\phi _{j_{0}k_{0},\alpha \alpha }=0,$ hence
varying $\{\phi _{j_{0}k_{0},\alpha \alpha }\}_{\alpha }$ in the domain $%
\{\phi _{jk,\alpha \alpha }\in R:\sum_{\alpha }\phi _{jk,\alpha \alpha }=0$,
$|\phi _{j_{0}k_{0},\alpha \alpha }|\leq \frac{1}{|B|}\}$, we get
\begin{eqnarray}
\sum_{\alpha ^{\prime }}(\Theta _{j_{0}k_{0},\alpha \alpha ,j^{\prime
}k^{\prime },\alpha ^{\prime }\alpha ^{\prime }}+\Theta _{j_{0}k_{0},\alpha
\alpha ,k^{\prime }j^{\prime },\alpha ^{\prime }\alpha ^{\prime }}^{\ast })
\end{eqnarray}
is independent of $\alpha .$

Again varying $\{\phi _{j_{0}k_{0},\alpha \alpha }\}_{\alpha }$ in the
domain $\{i\phi _{jk,\alpha \alpha }\in R:\sum_{\alpha }\phi _{jk,\alpha
\alpha }=0$, $|\phi _{j_{0}k_{0},\alpha \alpha }|\leq \frac{1}{|B|}\}$, we
get
\begin{eqnarray}
\sum_{\alpha ^{\prime }}(\Theta _{j_{0}k_{0},\alpha \alpha ,j^{\prime
}k^{\prime },\alpha ^{\prime }\alpha ^{\prime }}-\Theta _{j_{0}k_{0},\alpha
\alpha ,k^{\prime }j^{\prime },\alpha ^{\prime }\alpha ^{\prime }}^{\ast })
\end{eqnarray}
is independent of $\alpha .$

Thus
\begin{eqnarray}
\sum_{\alpha ^{\prime }}\Theta _{j_{0}k_{0},\alpha \alpha ,j^{\prime
}k^{\prime },\alpha ^{\prime }\alpha ^{\prime }}=\rho _{j^{\prime }k^{\prime
}}^{(j_{0}k_{0})}  \label{eqA25}
\end{eqnarray}
is independent of $\alpha .$

Step 3. Let $\phi _{jj,\alpha \alpha }=\frac{1}{|B|},\forall \ j,\alpha ,$
and $\phi _{j_{0}k_{0},\alpha _{0}\beta _{0}}=\phi _{k_{0}j_{0},\beta
_{0}\alpha _{0}}^{\ast }$ for fixed $j_{0},k_{0},\alpha _{0}\neq \beta _{0}$
and others $\phi _{jk,\alpha \beta }=0.$ We also let $|\phi
_{j_{0}k_{0},\alpha _{0}\beta _{0}}|\leq \frac{1}{|B|}$ to ensure $J_{\phi
}\geq 0.$ For this case, subtract \eqref{eqA16} from \eqref{eqA15} we get
\begin{eqnarray}
\phi _{j_{0}k_{0},\alpha _{0}\beta _{0}}\sum_{\alpha ^{\prime }}\Theta
_{j_{0}k_{0},\alpha _{0}\beta _{0},j^{\prime }k^{\prime },\alpha ^{\prime
}\alpha ^{\prime }} \ \ \ \ \ \   \notag \\
+\phi _{k_{0}j_{0},\beta _{0}\alpha _{0}}\sum_{\alpha ^{\prime }}\Theta
_{k_{0}j_{0},\beta _{0}\alpha _{0},j^{\prime }k^{\prime },\alpha ^{\prime
}\alpha ^{\prime }}=0.
\end{eqnarray}
Varying $\phi _{j_{0}k_{0},\alpha _{0}\beta _{0}}$ in the domain $|\phi
_{j_{0}k_{0},\alpha _{0}\beta _{0}}|\leq \frac{1}{|B|}$, we get
\begin{eqnarray}
\sum_{\alpha ^{\prime }}\Theta _{jk,\alpha _{0}\beta _{0},j^{\prime
}k^{\prime },\alpha ^{\prime }\alpha ^{\prime }}=0.  \label{eqA27}
\end{eqnarray}
Combine Eqs. \eqref{eqA17}, \eqref{eqA25}, \eqref{eqA27}, then we get Eq. %
\eqref{eq7}.

Step 4. Conversely, if Eqs. \eqref{eq7}, \eqref{eq8} hold, then for any $%
\phi \in \mathcal{C}_{AB}, $
\begin{eqnarray}
&&\sum_{\alpha ^{\prime }}[\Theta (\phi )]_{j^{\prime }k^{\prime },\alpha
^{\prime }\alpha ^{\prime }}  \notag \\
&=&\sum_{jk,\alpha \beta ,\alpha ^{\prime }}\phi _{jk,\alpha \beta }\Theta
_{jk,\alpha \beta ,j^{\prime }k^{\prime },\alpha ^{\prime }\alpha ^{\prime }}
\notag \\
&=&\sum_{jk,\alpha \beta }\phi _{jk,\alpha \beta }(\sum_{\alpha ^{\prime
}}\Theta _{jk,\alpha \beta ,j^{\prime }k^{\prime },\alpha ^{\prime }\alpha
^{\prime }})  \notag \\
&=&\sum_{jk,\alpha \beta }\phi _{jk,\alpha \beta }\delta _{\alpha \beta
}\rho _{j^{\prime }k^{\prime }}^{(jk)}  \notag \\
&=&\sum_{jk}(\sum_{\alpha }\phi _{jk,\alpha \alpha })\rho _{j^{\prime
}k^{\prime }}^{(jk)}  \notag \\
&=&\sum_{jk}\delta _{jk}\rho _{j^{\prime }k^{\prime }}^{(jk)} =\sum_{j}\rho
_{j^{\prime }k^{\prime }}^{(jj)}=\delta _{j^{\prime }k^{\prime }}.
\end{eqnarray}
Then we complete this proof.

\textbf{Corallory A1.} For $\Theta \in \mathcal{SC}_{ABA^{\prime }B^{\prime
}}$,
\begin{eqnarray}
trJ_{\Theta }=|B||A^{\prime }|.
\end{eqnarray}

\subsection{Proof of Proposition 2}

Note that $\Delta ^{A}\in \mathcal{C}_{AA},$ $\Delta ^{B}\in \mathcal{C}%
_{BB},$ and
\begin{eqnarray}
\Delta _{jk,\alpha \beta }^{A}=\delta _{jk}\delta _{j\alpha }\delta _{j\beta
}.
\end{eqnarray}
If $\phi \in \mathcal{IC}_{AB}$, then
\begin{eqnarray}
\phi (\rho ^{A})=\Delta ^{B}\phi \Delta ^{A}(\rho ^{A}),\forall \ \rho
^{A}\in \mathcal{D}_{A},
\end{eqnarray}
this yields
\begin{eqnarray}
\sum_{jk}\rho _{jk}^{A}\phi _{jk,\alpha \beta }=\delta _{\alpha \beta
}\sum_{j}\rho _{jj}^{A}\phi _{jj,\alpha \alpha }.  \label{33}
\end{eqnarray}
Let $\rho ^{A}=|j\rangle \langle j|,$ we get
\begin{eqnarray}
\phi _{jj,\alpha \beta }=\delta _{\alpha \beta }\phi _{jj,\alpha \alpha }.
\label{34}
\end{eqnarray}
Let $\rho ^{A}=\frac{I^{A}}{|A|}+\rho _{jk}^{A}|j\rangle \langle k|+\rho
_{kj}^{A}|k\rangle \langle j|,j<k,|\rho _{jk}^{A}|<\frac{1}{|A|},$ take it
into Eq. \eqref{33}, and together with Eq. \eqref{34}, we get
\begin{eqnarray}
\rho _{jk}^{A}\phi _{jk,\alpha \beta }+\rho _{kj}^{A}\phi _{kj,\alpha \beta
}=0.
\end{eqnarray}
Let $\rho _{jk}^{A}$ be real numbers we get $\phi _{jk,\alpha \beta }+\phi
_{kj,\alpha \beta }=0,$ let $\rho _{jk}^{A}$ be imaginary numbers we get $%
\phi _{jk,\alpha \beta }-\phi _{kj,\alpha \beta }=0$, then we get
\begin{eqnarray}
\phi _{jk,\alpha \beta }=\delta _{jk}\phi _{jj,\alpha \beta }.  \label{36}
\end{eqnarray}
Combining Eqs. \eqref{34} and \eqref{36} we then get Eq. \eqref{eq15} and
end this proof.

\subsection{Proof of Theorem 3}

Step 1. The first statement Eq. \eqref{eq16} is a result of Lemma 1 below.

\textbf{Lemma 1.} An $m\times n$ matrix $P=(P_{j\alpha })$ is called a row
stochastic matrix if it satisfies $P_{j\alpha }\geq 0$ for any $j,\alpha ,$
and $\sum_{\alpha =1}^{n}P_{j\alpha }=1$ for any $j.$ For any $m\times n$
row stochastic matrix $P,$ there exists a probability $\{p_{l}\}_{l=1}^{N},$
and row stochastic matrices $\{D^{(l)}\}_{l=1}^{N}$ such that each row of $%
D^{(l)}$ has just one nonzero entry ( that must be 1) for any $l,$ $%
P=\sum_{l=1}^{N}p_{l}D^{(l)},$ and $N\leq m(n-1)+1.$

Proof of Lemma 1. This proof is similar to the proofs of Theorem 2.1 and
Theorem 2.2 in Ref. \cite{Ye-2016} where for the case of $n\times n$. For
any $m\times n$ row stochastic matrix $P,$ denote $P^{(0)}=P,$ for $%
s=0,1,2,...,$ define $m\times n$ matrices $Q^{(s)}=(Q_{j\alpha }^{(s)})$ and
$P^{(s)}=(P_{j\alpha }^{(s)})$ as
\begin{eqnarray}
Q_{j\alpha }^{(s)}&=&\delta _{j,j_{s}}, \\
j_{s}&=&\max \{\alpha _{0}|P_{j,\alpha _{0}}^{(s)}\geq P_{j\alpha }^{(s)},
\forall \ \alpha \}, \\
\kappa (P^{(s)})&=&\min_{j}(\max_{\alpha }\{P_{j\alpha }^{(s)}\}), \\
P^{(s+1)}&=&P^{(s)}-\kappa (P^{(s)})Q^{(s)}.  \label{39}
\end{eqnarray}
For each iteration of $s$, $P^{(s+1)}$ has at least one zero entry more than
$P^{(s)}$. Hence after finite iterations we will attain $P^{(N)}=0.$ From
Eq. \eqref{39} and iterations we can reversely get an expression of $P$
which just has the form $P=\sum_{l=1}^{N}p_{l}D^{(l)}.$

Further since all $m\times n$ row stochastic matrices form a convex set with
dimensions $mn-m,$ then Carathodory's theorem yields that $N\leq m(n-1)+1.$

Step 2. We prove Eq. \eqref{eq17}. Note that, for $\phi \in \mathcal{C}_{AB}$
we can equivalently assume $|A|=|B|$ by adding extra dimensions to A or B
system. As a result, we restate the Proposition 1 in Ref. \cite%
{ChitambarGour-2016-PRL} as follows.

\textbf{Lemma 2.} A channel $\phi \in \mathcal{C}_{AB}$ is a PIO iff it can
be expressed as a convex combination of channels each having Kraus operators
$\{M_{n}\}_{n}$ of the form
\begin{eqnarray}
M_{n}=U_{n}P_{n}=\sum_{j=1}^{|A|}e^{i\theta _{j}}|f_{n}(j)\rangle \langle
j|P_{n}
\end{eqnarray}
where $\{P_{n}\}_{n}$ form an orthogonal and complete set of incoherent
projectors on system $A$ and $f_{n}(j)\in \{\alpha \}_{\alpha =1}^{|B|},$ $%
f_{n}(j)\neq f_{n}(k)$ for $j\neq k.$

Now, let $\phi =\{M_{n}\}_{n=1}^{|A|}\in \mathcal{C}_{AB}$ as
\begin{eqnarray}
P_{n}=|n\rangle \langle n|,
\end{eqnarray}
then
\begin{eqnarray}
M_{n}&=&e^{i\theta _{n}}|f_{n}(n)\rangle \langle n|, \\
|M_{n}\rangle &=&e^{i\theta _{n}}|n\rangle |f_{n}(n)\rangle , \\
|M_{n}\rangle \langle M_{n}|&=&|n\rangle \langle n|\otimes |f_{n}(n)\rangle
\langle f_{n}(n)|, \\
J_{\phi }&=&\sum_{n}|n\rangle \langle n|\otimes |f_{n}(n)\rangle \langle
f_{n}(n)|,
\end{eqnarray}
which just is an IC. Together with Eq. \eqref{eq16}, we get $\mathcal{IC}%
_{AB}\subset \mathcal{PIO}_{AB}.$

To show $\mathcal{IC}_{AB}\neq \mathcal{PIO}_{AB},$ let channel $\phi
=\{U\}\in \mathcal{C}_{AB}$ as $U=\sum_{j=1}^{|A|}|f(j)\rangle \langle j|$
with $f(j)\in \{\alpha \}_{\alpha =1}^{|B|},$ $f(j)\neq f(k)$ for $j\neq k,$
then
\begin{eqnarray}
|U\rangle &=&\sum_{j=1}^{|A|}|j\rangle |f(j)\rangle , \\
J_{\phi }&=&|U\rangle \langle U|=\sum_{jk}|j\rangle \langle k|\otimes
|f(j)\rangle \langle f(k)|,
\end{eqnarray}
which is evidently not an IC for $|A|\geq 2.$

\subsection{Proof of Theorem 4}

Step 1. Suppose $\Theta =\{\mathcal{U}\}\in \mathcal{SC}_{ABA^{\prime
}B^{\prime }}$ is preunitary, then for any $\phi \in \mathcal{C}_{AB},$
\begin{eqnarray}
\Theta _{jk,\alpha \beta ,j^{\prime }k^{\prime },\alpha ^{\prime }\beta
^{\prime }}=\mathcal{U}_{j^{\prime }j,\alpha ^{\prime }\alpha }\mathcal{U}%
_{k^{\prime }k,\beta ^{\prime }\beta }^{\ast }, \ \ \ \ \ \ \ \ \ \ \ \ \ \
\ \ \ \ \ \  \\
\sum_{\alpha ^{\prime }}\Theta _{jk,\alpha \beta ,j^{\prime }k^{\prime
},\alpha ^{\prime }\alpha ^{\prime }} =\sum_{\alpha ^{\prime }}\mathcal{U}%
_{j^{\prime }j,\alpha ^{\prime }\alpha }\mathcal{U}_{k^{\prime }k,\alpha
^{\prime }\beta }^{\ast }=\delta _{\alpha \beta }\rho _{j^{\prime }k^{\prime
}}^{(jk)},  \notag \\
\\
\sum_{j}\rho _{j^{\prime }k^{\prime }}^{(jj)} =\delta _{j^{\prime }k^{\prime
}}. \ \ \ \ \ \ \ \ \ \ \ \ \ \ \ \ \ \ \ \ \ \ \ \ \ \ \ \ \ \ \ \ \ \
\end{eqnarray}
For fixed $j^{\prime} =k^{\prime }=j=k=1,$ $\{\mathcal{U}_{11,\alpha
^{\prime }\alpha }\}_{\alpha ^{\prime }\alpha }$ satisfies $\sum_{\alpha
^{\prime }}\mathcal{U}_{11,\alpha ^{\prime }\alpha }\mathcal{U}_{11,\alpha
^{\prime }\beta }^{\ast }=\delta _{\alpha \beta }\rho _{11}^{(11)},$ then
\begin{equation}
\{\mathcal{M}_{11,\alpha ^{\prime }\alpha }\}_{\alpha ^{\prime }\alpha }=%
\sqrt{P_{11}}V,
\end{equation}
where $P_{11}\geq 0$, $V$ is a $|B^{\prime }|\times |B|$ isometry, i.e. $%
V^{\dagger }V=I^{B}$. Note that for $P_{11}>0,$ $V^{\dagger }V=I^{B}$
requires $|B|\leq |B^{\prime }|.$

\textbf{Lemma 3.} If $\{|\varphi _{i}\rangle \}_{i=1}^{n}$ and $\{|\psi
_{i}\rangle \}_{i=1}^{n}$ are two orthonormal bases of Hilbert space $H$,
and for $\forall $ $j,k,\langle \varphi _{j}|\psi _{k}\rangle =\delta
_{jk}\langle \varphi _{j}|\psi _{j}\rangle ,\langle \varphi _{j}|\psi
_{j}\rangle =\langle \varphi _{k}|\psi _{k}\rangle ,$ then $|\psi
_{j}\rangle =e^{i\theta }|\varphi _{j}\rangle $ for $\forall $ $j$ with real
number $\theta .$

Proof hint: expand $\{|\psi _{j}\rangle \}_{j=1}^{n}$ in $\{|\varphi
_{k}\rangle \}_{k=1}^{n}.$

Using this lemma to $\{\mathcal{U}_{j^{\prime }j,\alpha ^{\prime }\alpha
}\}_{\alpha ^{\prime }\alpha }$ and $\{\mathcal{U}_{11,\alpha ^{\prime
}\alpha }\}_{\alpha ^{\prime }\alpha }$, we get
\begin{eqnarray}
\{\mathcal{U}_{j^{\prime }j,\alpha ^{\prime }\alpha }\} =\sqrt{P_{j^{\prime
}j}}e^{i\theta _{j^{\prime }j}}V
\end{eqnarray}
with $P_{j^{\prime }j}\geq 0$, $\theta _{j^{\prime }j}\in R.$ That is
\begin{eqnarray}
\mathcal{U=}U\otimes V,
\end{eqnarray}
with $U_{j^{\prime }j}=\sqrt{P_{j^{\prime }j}}e^{i\theta _{j^{\prime }j}}.$ $%
\sum_{j}\rho _{j^{\prime }k^{\prime }}^{(jj)}=\delta _{j^{\prime }k^{\prime
}}$ yields that $U$ is a $|A^{\prime }|\times |A|$ coisometry, i.e. $%
UU^{\dagger }=I^{A^{\prime }}.$ Again, $UU^{\dagger }=I^{A^{\prime }}$
requires $|A|\geq |A^{\prime }|.$

We can write $\mathcal{U=}U\otimes V$ as
\begin{eqnarray}
\mathcal{U}_{j^{\prime }j,\alpha ^{\prime }\alpha }=U_{j^{\prime
}j}V_{\alpha ^{\prime }\alpha }.
\end{eqnarray}

Step 2. Now suppose $\Theta =\{\mathcal{U\}}\in \mathcal{ISC}_{AB}.$ For $%
\phi \in \mathcal{IC}_{AB}$ as
\begin{eqnarray}
\phi _{jj,f(j)f(j)}=1, \forall \ j,
\end{eqnarray}
with $f(j)\in \{\alpha \}_{\alpha =1}^{|B|},$ and other elements $\phi
_{jk,\alpha \beta }=0.$ For this case
\begin{eqnarray}
&&[\Theta (\phi )]_{j^{\prime }k^{\prime },\alpha ^{\prime }\beta ^{\prime }}
\notag \\
&=&\sum_{j}\phi _{jj,f(j)f(j)}\Theta _{jj,f(j)f(j),j^{\prime }k^{\prime
},\alpha ^{\prime }\beta ^{\prime }}  \notag \\
&=&\sum_{j}\Theta _{jj,f(j)f(j),j^{\prime }k^{\prime },\alpha ^{\prime
}\beta ^{\prime }}  \notag \\
&=&\sum_{j}\mathcal{U}_{j^{\prime }j,\alpha ^{\prime }f(j)}\mathcal{U}%
_{k^{\prime }j,\beta ^{\prime }f(j)}^{\ast }.  \notag \\
&=&\sum_{j}U_{j^{\prime }j}U_{k^{\prime }j}^{\ast }V_{\alpha ^{\prime
}f(j)}V_{\beta ^{\prime }f(j)}^{\ast }.
\end{eqnarray}
Let $j^{\prime } =k^{\prime } $, $\alpha ^{\prime }\neq \beta ^{\prime
},f(j)=\alpha $ for $\forall \ j,$
\begin{eqnarray}
0=[\Theta (\phi )]_{j^{\prime }j^{\prime },\alpha ^{\prime }\beta ^{\prime
}}=(\sum_{j}U_{j^{\prime }j}U_{j^{\prime }j}^{\ast })V_{\alpha ^{\prime
}\alpha }V_{\beta ^{\prime }\alpha }^{\ast }  \notag \\
=V_{\alpha ^{\prime }\alpha }V_{\beta ^{\prime }\alpha }^{\ast },
\end{eqnarray}
as a result, each column of $V$ has at most one nonzero element. And since $%
V^{\dagger }V=I^{B}$, then $V$ has just one nonzero element with modulus 1
in each column and has at most one nonzero with modulus 1 in each row.

Let $j^{\prime } \neq k^{\prime },$ suppose $V_{\alpha ^{\prime }\alpha
}\neq 0,\ $then $V_{\alpha ^{\prime }\beta }=0$ for $\beta \neq \alpha ,$
let $\alpha ^{\prime }=\beta ^{\prime },$ $f(1)=\alpha ,$ $f(j)\neq \alpha $
for $j\neq 1,$
\begin{eqnarray}
0=[\Theta (\phi )]_{j^{\prime }k^{\prime },\alpha ^{\prime }\alpha ^{\prime
}}=\sum_{j}U_{j^{\prime }j}U_{k^{\prime }j}^{\ast }V_{\alpha ^{\prime
}f(j)}V_{\alpha ^{\prime }f(j)}^{\ast }  \notag \\
=U_{j^{\prime }1}U_{k^{\prime }1}^{\ast },
\end{eqnarray}
this yields that the first column of $U$ has at most one nonzero element.
Similarly every column of $U$ has at most one nonzero element.

\subsection{Proof of Proposition 8}

Step 1. We show that the superchannel $\Theta $ with $\Theta (\phi )=\chi
\circ \phi $ is an ISC. Let $A=A^{\prime },$ then $\Theta \in \mathcal{SC}%
_{ABA^{\prime }B^{\prime }}.$ From the convexity of coherence for channels
in (C3) and the structure of $\mathcal{IC}$ in Eq. \eqref{eq16}, we only need to
consider the case that $\chi _{\alpha \beta ,\alpha ^{\prime }\beta ^{\prime
}}=\delta _{\alpha \beta }\delta _{\alpha ^{\prime }\beta ^{\prime }}\delta
_{\alpha ^{\prime },f(\alpha )}$ with $f(\alpha )\in \{\alpha ^{\prime
}\}_{\alpha ^{\prime }=1}^{|B^{\prime }|}.$ It follows that
\begin{eqnarray}
&&[\Theta (\phi )]_{j^{\prime }k^{\prime },\alpha ^{\prime }\beta ^{\prime }}
\notag \\
&=&\sum_{jk\alpha \beta }\phi _{jk,\alpha \beta }\Theta _{jk,\alpha \beta
,j^{\prime }k^{\prime },\alpha ^{\prime }\beta ^{\prime }}  \notag \\
&=&[\chi \circ \phi ]_{j^{\prime }k^{\prime },\alpha ^{\prime }\beta
^{\prime }}  \notag \\
&=&\sum_{\alpha \beta }\phi _{j^{\prime }k^{\prime },\alpha \beta }\chi
_{\alpha \beta ,\alpha ^{\prime }\beta ^{\prime }}  \notag \\
&=&\sum_{\alpha }\phi _{j^{\prime }k^{\prime },\alpha \alpha }\delta
_{\alpha ^{\prime }\beta ^{\prime }}\delta _{\alpha ^{\prime },f(\alpha )},
\end{eqnarray}%
consequently,
\begin{eqnarray}
\Theta _{jk,\alpha \beta ,j^{\prime }k^{\prime },\alpha ^{\prime }\beta
^{\prime }}=\delta _{jj^{\prime }}\delta _{kk^{\prime }}\delta _{\alpha
\beta }\delta _{\alpha ^{\prime }\beta ^{\prime }}\delta _{\alpha ^{\prime
},f(\alpha )},
\end{eqnarray}
which admits the decomposition of Eq. \eqref{eq10} with
\begin{eqnarray}
\mathcal{M}_{m,j^{\prime }j,\alpha ^{\prime }\alpha }=\delta _{m\alpha
}\delta _{jj^{\prime }}\delta _{\alpha ^{\prime },f(\alpha )}, \ \ \ \ \ \ \
\ \ \ \ \ \  \\
\mathcal{M}_{m}=\sum_{jj^{\prime }\alpha\alpha^{\prime }}\delta _{m\alpha
}\delta _{jj^{\prime }}\delta _{\alpha ^{\prime },f(\alpha )}|j^{\prime
}\rangle |\alpha ^{\prime }\rangle \langle j|\langle \alpha |.
\end{eqnarray}%
Compare to Eq. \eqref{eq18} we see that $\Theta =\{\mathcal{M}_{m}\}_{m}\in
\mathcal{ISC}_{ABA^{\prime }B^{\prime }}.$

Step 2. We show that $\Theta (\phi )=\phi \circ \chi $ is an ISC. Let $%
B=B^{\prime },$ then $\Theta \in \mathcal{SC}_{ABA^{\prime }B^{\prime }}.$
From Eq. \eqref{eq16}, for $|A^{\prime }|\leq |A|$, we have
\begin{eqnarray}
\{\chi \in \mathcal{C}_{A^{\prime }A}|\chi _{j^{\prime }k^{\prime
},jk}=\delta _{j^{\prime }k^{\prime }}\delta _{jk}\phi _{j^{\prime
}j^{\prime },jj}, \nonumber \\
\sum_{j^{\prime }}\chi _{j^{\prime }j^{\prime },jj}\leq
1,\forall \ j\}  \notag \\
=conv\{\chi \in \mathcal{C}_{A^{\prime }A}|\chi _{j^{\prime }k^{\prime
},jk}=\delta _{j^{\prime }k^{\prime }}\delta _{jk}\delta _{j,f(j^{\prime
})},  \notag \\
f(j^{\prime }) \neq f(k^{\prime })\text{\ for \ }j^{\prime }\neq k^{\prime
}\},
\end{eqnarray}%
where $f(j^{\prime })\in \{j\}_{j=1}^{|A|}.$ Thus we only need to consider
the case that $\chi \in \mathcal{C}_{A^{\prime }A},$ $\chi _{j^{\prime
}k^{\prime },jk}=\delta _{j^{\prime }k^{\prime }}\delta _{jk}\delta
_{j,f(j^{\prime })},$ $f(j^{\prime })\neq f(k^{\prime })$ for $j^{\prime
}\neq k^{\prime }.$ Similar to Eqs. (A60-A62) we get
\begin{eqnarray}
\Theta _{jk,\alpha \beta ,j^{\prime }k^{\prime },\alpha ^{\prime }\beta
^{\prime }}=\delta _{jk}\delta _{j^{\prime }k^{\prime }}\delta _{\alpha
\alpha ^{\prime }}\delta _{\beta \beta ^{\prime }}\delta _{j,f(j^{\prime })},
\\
\mathcal{M}_{m,j^{\prime }j,\alpha ^{\prime }\alpha }=\delta _{mj^{\prime
}}\delta _{\alpha \alpha ^{\prime }}\delta _{j,f(j^{\prime })},\ \ \ \ \ \ \
\ \ \ \ \ \  \\
\mathcal{M}_{m} =\sum_{jj^{\prime }\alpha \alpha ^{\prime }}\delta
_{mj^{\prime }}\delta _{\alpha \alpha ^{\prime }}\delta _{j,f(j^{\prime
})}|j^{\prime }\rangle |\alpha ^{\prime }\rangle \langle j|\langle \alpha |.
\end{eqnarray}
Comparing to Eq. \eqref{eq18} we see that $\Theta =\{\mathcal{M}_{m}\}_{m}\in
\mathcal{ISC}_{ABA^{\prime }B^{\prime }}.$

\bibliographystyle{apsrev4-1}
\bibliography{refcoherenceofchannels}

\end{document}